\documentclass[conference]{IEEEtran}

\pdfoutput=1

\IEEEoverridecommandlockouts

\usepackage{cite}

\ifCLASSINFOpdf
  \usepackage[pdftex]{graphicx}
\else
\fi
\usepackage[cmex10]{amsmath}
\usepackage{amsfonts}
\usepackage{amssymb}
\usepackage{algorithmic}
\ifCLASSOPTIONcompsoc
  \usepackage[caption=false,font=normalsize,labelfont=sf,textfont=sf]{subfig}
\else
  \usepackage[caption=false,font=footnotesize]{subfig}
\fi
\usepackage{url}

\usepackage{tikz}

\newcommand*\mycirc[1]{%
  \begin{tikzpicture}
    \node[draw,circle,inner sep=1pt] {#1};
  \end{tikzpicture}}

\usepackage{color}  
  
\definecolor{gris}{gray}{0.25}

\begin{document}
%
\title{Towards Statistical Prioritization for Software Product Lines Testing}

\author{
    \IEEEauthorblockN{Xavier Devroey, Maxime Cordy, \\ Gilles Perrouin\textsuperscript{$^\diamondsuit$}\thanks{$^\diamondsuit$FNRS Postdoctoral Researcher}, \\Pierre-Yves Schobbens}
    \IEEEauthorblockA{PReCISE Research Center, \\Faculty of Computer Science, \\ University of Namur, Belgium \\ \{xde,mcr,gpe,pys\}@info.fundp.ac.be}
\and
    \IEEEauthorblockN{Axel Legay}
    \IEEEauthorblockA{INRIA Rennes Bretagne Atlantique, \\ France \\ axel.legay@inria.fr}
\and
    \IEEEauthorblockN{Patrick Heymans}
    \IEEEauthorblockA{PReCISE Research Center, \\ Faculty of Computer Science, \\ University of Namur, Belgium \\ phe@info.fundp.ac.be}
    \IEEEauthorblockA{INRIA Lille-Nord Europe, \\ Universit\'e Lille 1 -- LIFL -- CNRS , France}
}


%


\IEEEspecialpapernotice{Extended version published at VaMoS '14 (\url{http://dx.doi.org/10.1145/2556624.2556635)}}

\maketitle

\begin{abstract}
Software Product Lines (SPL) are inherently difficult to test due to the combinatorial explosion of the number of products to consider. To reduce the number of products to test, sampling techniques such as combinatorial interaction testing have been proposed.  They usually start from a feature model and apply a coverage criterion (e.g. pairwise feature interaction or dissimilarity) to generate tractable, fault-finding, lists of configurations to be tested. Prioritization can also be used to sort/generate such lists, optimizing coverage criteria or weights assigned to features.  However, current sampling/prioritization techniques barely take product behavior into account. We explore how ideas of statistical testing, based on a usage model (a Markov chain), can be used to extract configurations of interest according to the likelihood of their executions.  These executions are gathered in featured transition systems, compact representation of SPL behavior.  We discuss possible scenarios and give a prioritization procedure illustrated on an example.   
\end{abstract}


%
\IEEEpeerreviewmaketitle

\section{Introduction}

Software Product Line (SPL) engineering is based on the idea that products of the same family may be built by systematically reusing assets, some of them being common to all members whereas others are only shared by a subset of the family. Such variability is commonly captured  by the notion of \emph{feature}, i.e., an unit of difference between products. A product member of the SPL is  a valid combination of features. Individual features can be specified using  languages such as UML, while their inter-relationships are organized in a \emph{Feature Diagram} (FD) \cite{kang_feature-oriented_1990}. An FD thus (abstractly) describes all valid combinations of features (called \emph{configurations} of the FD), that is all the products of the SPL.  

In this paper, we are interested in SPL testing. As opposed to to classical testing approaches, where the testing process only considers one software product, SPL testing is concerned about how to minimize the test effort related to a given the SPL (i.e., all the valid products of the SPL). The size of this set is roughly equal to $2^N$, where $N$ represents the number of features of the SPL. This number may vary from about 10 ($2^{10}$ possible products) in small SPLs to thousands of features ($2^{1000}$ possible products)  in complex systems such as the Linux kernel. Automated Model-Based Testing  \cite{utting2007practical} and shared execution \cite{DBLP:conf/issre/KimKB12}, where tests can be reused amongst products, are candidates to reduce such effort. 

Still, testing all products of the SPL is only possible for small SPLs, given their exponential growth with the number of features. Hence, one of the main questions arising in such a situation is: How to extract and prioritize relevant products? Existing approaches consider sampling products by using a coverage criterion on the FD (such as all valid 2-tuples of features:  pairwise \cite{PerrouinOSKBT12,Cohen2007}) and rank products with respect with respect to coverage satisfaction (e.g. the number of tuples covered). An alternative is to label each feature with weights and prioritize configurations accordingly \cite{splc13,DBLP:conf/models/JohansenHFES12}. These methods actually help testers to scope more finely and flexibly relevant products to test than using a covering criteria alone. Yet, these approaches only sample products based on the FD which does not account for product behavior:  they are just configurations of the FD. Furthermore, assigning meaningful weights on thousands of features can be tricky if no other information is available. 
         
Statistical testing \cite{Whittaker:1994fk} proposes to generate test cases based on a \emph{usage model} represented by a \emph{Discrete Time Markov Chain} (DTMC). This usage model represents the usage scenarios of the software as well as their probability.  This allows one to determine the relative importance of execution scenarios (with respect to other). This paper explores the possibility  of using statistical testing to sample and prioritize products of an SPL. The basic idea is to  focus on ``most probable'' (respectively ``less probable'' products), i.e. products that are able to execute highly probable (respectively.  improbable) traces of the DTMC. Since black box usage scenarios may not relate directly to features, we propose to use  a compact representation of SPL behaviour, \emph{Featured Transition Systems} (FTSs) to determine which traces are legal with respect to the SPL and associate related products. In fact, we construct another FTS, which maps to the selected traces. This FTS represent the behavior of the set of products of interest and is amenable to various testing and verification techniques. 

The remainder of this paper is organized as follows: Section \ref{background} presents the theoretical background underlying our vision presented in section \ref{approach}. Section \ref{relatedWork} discussed related research and Section \ref{conclusion} concludes the paper with challenges and future research directions.

\section{Background} \label{background}

\begin{figure}[t]
\centering
\subfloat[Feature Diagram (FD)]{
    \includegraphics[width=0.45\textwidth]{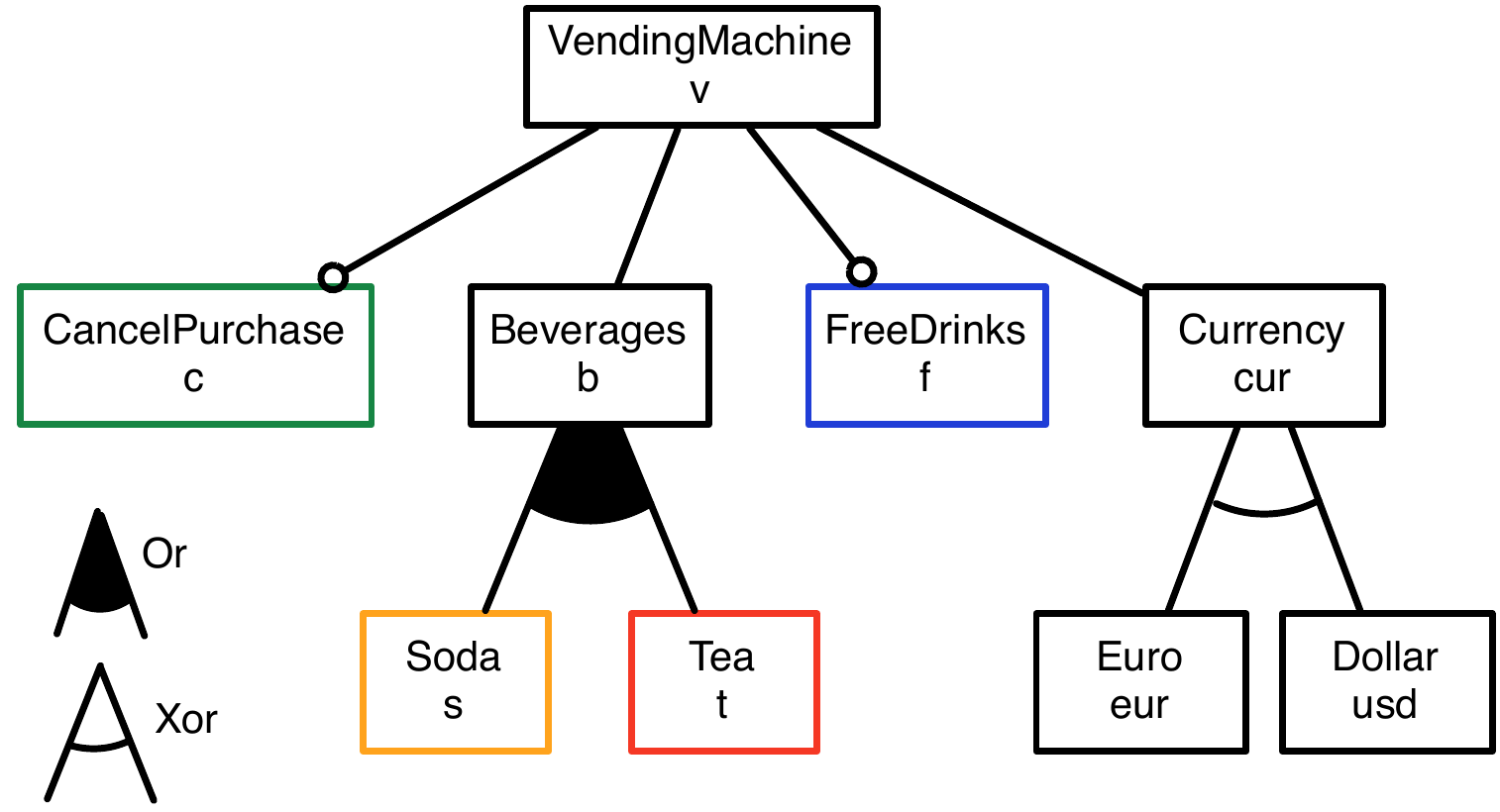}
    \label{fig_FD_vending_machine}
}
\hfil
\subfloat[Featured Transition System (FTS)]{
	\includegraphics[width=0.45\textwidth]{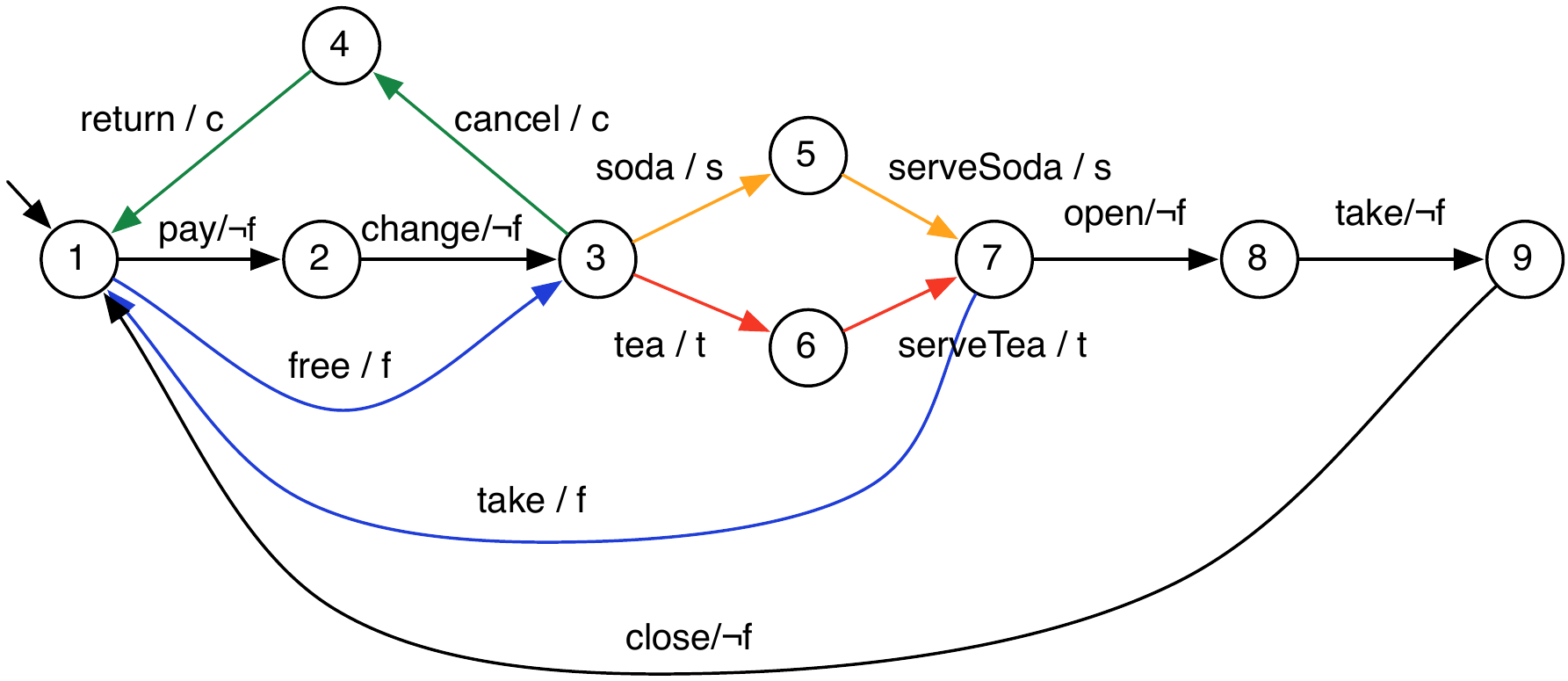}
	\label{fig_FTS_vending_machine}
}
\hfil
\subfloat[Usage model (DTMC)]{
    \includegraphics[width=0.45\textwidth]{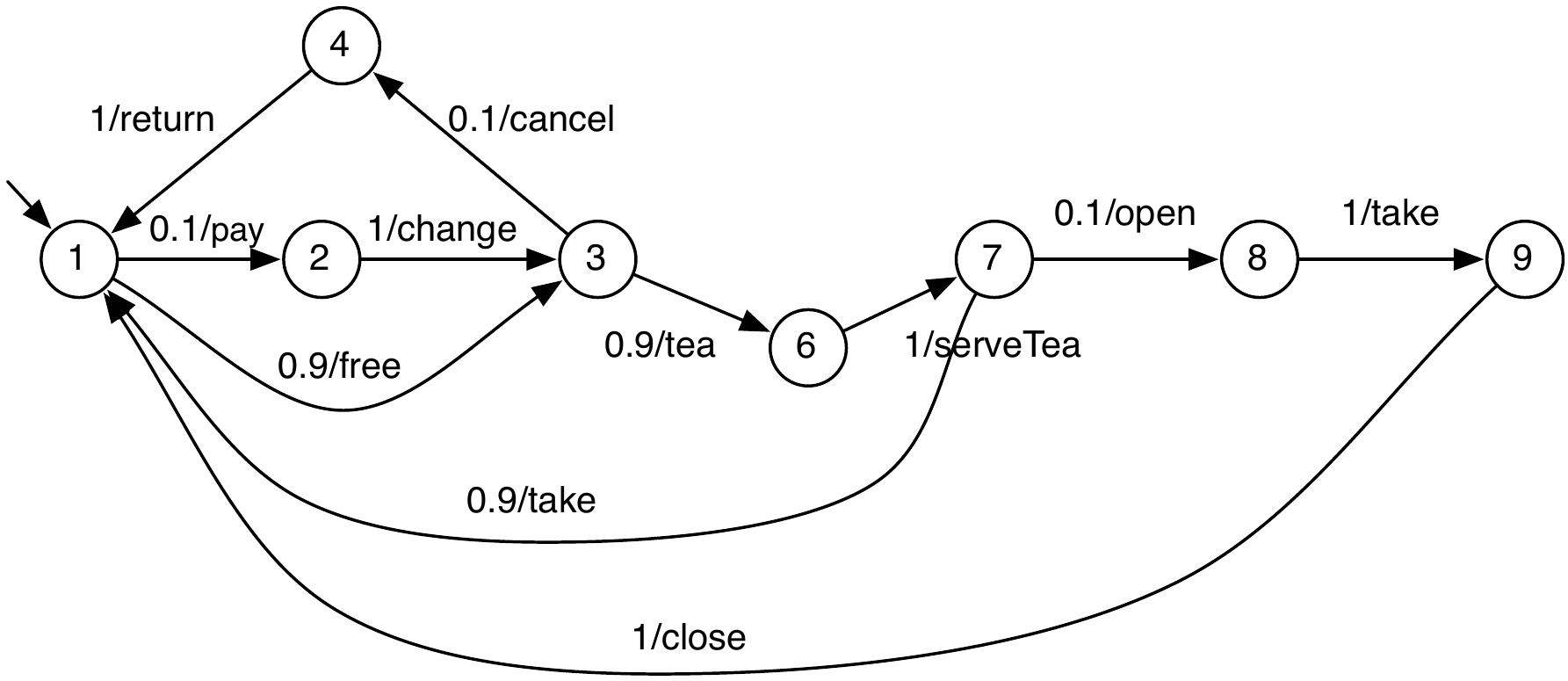}
    \label{fig_DTMC_vending_machine}
}
\caption{The soda vending machine example \cite{Classen2013}}
\label{fig_model_used}
\end{figure}

In this section, we present the foundations underlying our approach:  \emph{SPL modeling} and \emph{statistical testing}. 
\subsection{SPL Modelling} 


A key concern in SPL modeling is how to represent variability. To achieve this purpose, SPL engineers usually reason in terms of features. In \cite{pohl2005software} Pohl et al. define features as end-user visible characteristics of a system. Relations and constraints between features are usually represented in a Feature Diagram (FD)~\cite{kang_feature-oriented_1990}. For example, Fig. \ref{fig_FD_vending_machine} presents the FD of a soda vending machine \cite{Classen2013}. A product derived from this diagram will correspond to a set of selected features. Here, $\{v, b, t, cur, usd \}$ corresponds to a machine that sells only tea and accept only dollar. FDs have been equipped with formal semantics \cite{Schobbens2007}, automated analyses and tools \cite{kang_feature-oriented_1990} for more than 20 years. A common semantics associated to a FD $d$ (noted $[\![d]\!]$) is the set of all the valid products allowed by $d$.

\subsubsection{Behavioural Modelling}


Different formalisms may be used to model the behavior of a system. To allow the explicit mapping from feature to SPL behavior, Featured Transition Systems (FTSs) \cite{Classen2013} were proposed. FTSs are Transition Systems (TSs) where each transition is labelled with a feature expression (i.e., a boolean expression over features of the SPL), specifying for a given FD in which products the transition may be fired.  Thus it is possible to determine products that are the cause of a violation or a failed test. Formally, an FTS is a tuple $(S, Act, trans, i, d, \gamma)$ where:
\begin{itemize}
\item $(S, Act, trans, i)$ is a classical TS with $S$ a set of states, $Act$ a set of actions, $trans \subseteq S \times Act \times S $ a set of transitions (where $(s_1, \alpha, s_2) \in trans$ is sometimes noted $s_1 \overset{\alpha}{\longrightarrow} s_2 $) and $i \in S$  an initial state;
\item $d$ is a FD;
\item  $\gamma : trans \rightarrow [\![d]\!] \rightarrow \mathbb{B}$ is a total function, labeling each transition with a function that associates to each valid product a boolean expression indicating if it may fire the transition or not. This function is noted using a feature expression (which is basically a boolean expression with features' name as variables) meaning that only products that satisfy this feature expression may fire the transition. For instance: $\neg f$ in Fig. \ref{fig_FTS_vending_machine} indicates that only products that have not the $free$ feature may fire the $pay, change, open, take$ and $close$ transitions.
\end{itemize}

The semantics of a given FTS is a function that associates each valid product with its set of finite and infinite executions, i.e. all the possible paths in the graph starting from the initial state available for this specific product. According to this definition, an FTS is actually a behavioral model of a whole SPL. Fig. \ref{fig_FTS_vending_machine} presents the FTS modeling a vending machine SPL. For instance, transition $\mycirc{3} \overset{pay / \neg f}{\longrightarrow}  \mycirc{4}$ is labelled with the feature expression $c$. This means that only the products that do have the feature $Cancel\ (c)$ are able to execute the transition. This definition differs from the one presented in \cite{Classen2013}, where only infinite paths are considered. In a testing context, one may also be interested in finite paths.



\subsection{Statistical Testing} 

Whittaker and Thomason introduce the notion of usage model in \cite{Whittaker:1994fk} and define it as a TS where transitions are associated to probabilities. A probability $p_i$ on a transition $t_i = (s_i, \alpha, s_j)$represents the fact that if the system is in  state $s_i$, the transition $t_i$ has $p_i$ chances to be fired. Formally, a usage model will be represented by a DTMC, which is a tuple $(S, Act, trans, P, \tau)$ where :
\begin{itemize}
\item $(S, Act, trans)$ is a TS;
\item $P : S \times S \rightarrow [0,1] $ is the probability matrix which gives for two states $( s_i , s_j )$ the probability for the system in state $s_i$ to go in the state $s_j$;
\item $ \tau : S \rightarrow [0,1] $ is the vector containing the probabilities to be in the initial state when the system starts, with the following constraint : $\exists i : (\tau (i) = 1 \wedge \forall j \neq i : \tau (j) = 0)$;
\item $\forall s_i \in S : \sum_{s_j \in S} P( s_i , s_j ) = 1$ the total of the probabilities of the transitions leaving a state must be equal to 1.
\end{itemize}
Actions are just labels used to annotate transitions without changing the semantics of the DTMC. In our case, they are used to relate traces of the DTMC with executions of the FTS.

\section{Approach} \label{approach}


In our approach, we consider three models: a FD $d$ to represent the features and their constraints (in Fig. \ref{fig_FD_vending_machine}), an FTS $fts$ to represent the behaviour of the SPL (in Fig. \ref{fig_FTS_vending_machine}) and a usage model represented by a DTMC $dtmc$ (in Fig. \ref{fig_DTMC_vending_machine}) with the following constraints:
\begin{itemize}
\item The feature diagram of $fts$ is $d$;
\item States, actions and transitions of $dtmc$ are included in the states, actions and transitions (respectively) of $fts$: $S_{dtmc} \subseteq S_{fts} \wedge Act_{dtmc} \subseteq Act_{fts} \wedge trans_{dtmc} \subseteq trans_{fts}$;
\item The initial state of $fts$, $i$ has a probability of $1$ to be executed first in $dtmc$ : $\tau (i) = 1$
\end{itemize}

We deliberately chose not to integrate the DMTC with the FTS in a single model. This separation of concerns is motivated as follows:

\begin{itemize}
\item We may want to integrate the approach with existing software which does not take variability into account such as MaTeLO, a MBT tool which uses DTMC as input model \footnote{see: \url{http://all4tec.net/index.php/en/model-based-testing/20-markov-test-logic-matelo}};

\item The DTMC can be obtained from either users trying the software under test, extracted from logs, or from running code. These extractions methods are agnostic of the features of the system they are applied to;

\item Since the DTMC is built from existing software executions, it may be incomplete (as in Fig. \ref{fig_DTMC_vending_machine}). Some products (or subsets of their behaviors) may simply not be exercised in the usage model resulting in missing transitions in the DTMC. Keeping the FTS and usage models separate is helpful to identify and correct such issues.  
\end{itemize}
 
The fact that a usage model is created from partial (i.e., finite) observations of the products without consideration of their features allows paths in the DTMC that are inconsistent for the SPL. For example in the usage model of Fig. \ref{fig_DTMC_vending_machine}, one can follow the path $pay$, $change$, $tea$, $serveTea$, $take$. This path actually mixes ``pay machine'' (feature $f$ not enabled)  and ``free machine'' (feature $f$ enabled). Since the DTMC is never used alone, such situations are easy to spot using the FTS. 

There are now two possible testing scenarios: product based test derivation (top-down)  and family based test prioritization (bottom-up). The classification product/family based comes from \cite{VonRhein2013}. 

\subsection{Product Based Test Derivation} 

Product based test derivation is straightforward: one selects one product (by selecting features in the FD), projects it onto the FTS, giving a TS with only the transitions of the product, prunes the DTMC to keep the following property true : $S_{dtmc} \subseteq S_{ts} \wedge Act_{dtmc} \subseteq Act_{ts} \wedge trans_{dtmc} \subseteq trans_{ts}$. Probabilities of the removed transitions need to be distributed on siblings (since the property $\forall s_i \in S : \sum_{s_j \in S} P( s_i , s_j ) = 1$ has to hold). Finally, we generate test cases using classical statistical testing algorithms on the DTMC~\cite{Whittaker:1994fk,feliachi2010generating}.  A similar testing process is proposed by Samih and Baudry~\cite{Samih2012}.  Product selection is made on an orthogonal variability model (OVM) and mapping between the OVM and the DTMC  (implemented using MaTeLo) is provided via explicit traceability links to functional requirements. This process thus requires to perform selection of products of interest on the variability model and does not exploit probabilities and traces of the DTMC during such selection. Additionally, they assume that tests for all products of the SPL are modeled in the DTMC. This assumption may be too strong in certain cases and delay actual testing since designing the complete DTMC for a large SPL may take time. We thus explore a scenario where the DTMC drives product selection and prioritization.  


\subsection{Family Based Test Prioritization} 

\begin{figure*}[t]
\centering
\includegraphics[width=0.88\textwidth]{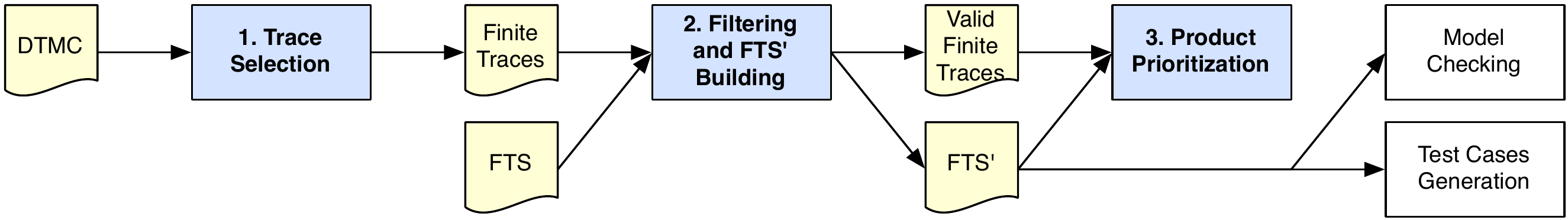}
\caption{Family based test prioritization approach}
\label{fig_family_based_approach}
\end{figure*}

Contrary to product based test derivation, our approach (in Fig. \ref{fig_family_based_approach}) only assumes partial coverage of the SPL by the usage model. For instance, the DTMC represented in Fig. \ref{fig_DTMC_vending_machine} does not cover serving soda behavior because no user/tester exercised it. The key idea is to generate sequences of actions (i.e., finite traces) from the DTMC according to their probability to happen (step 1). For example, one may be interested in analyzing behaviors including serving teas for free, which will correspond to the sequence of actions $(free, tea,serveTea,take)$, since it is highly probable ($p=0.729$). On the contrary, one may be interested in low probability because it can mean poorly tested or irrelevant products. 

The generated sequences are filtered using the FTS in order to keep only sequences that may be executed by at least one product of the SPL (step~2). The result will be a FTS', corresponding to a pruned FTS according to the extracted sequences. Each valid sequence of actions is combined with the FTS' to generated a set of products that may effectively execute this sequence. The probability of the sequence to be executed allows to prioritize products exercising the behavior described in the FTS' (step~3).

\subsubsection{Trace Selection in the DTMC}

The first step is to extract sequences of actions (i.e., finite traces) from the DTMC according to desired parameters provided by the tester. We define a finite trace as a finite path (i.e., a finite sequence of labels) in a TS. This may differ from the standard notion of trace but since we are in our context, infinite traces are not really useful. 
Formally, a finite trace $t$ corresponds to a tuple of labels : $t=(\alpha_1 , ... , \alpha_n)$ such as  $ \exists s_i, s_j \in S_{TS} \wedge s_i \overset{(\alpha_1 , ... , \alpha_n)}{\Longrightarrow} s_j $ where $s_i \overset{(\alpha_1 , ... , \alpha_n)}{\Longrightarrow} s_j$ denotes the existence of a path ($s_i \overset{\alpha_1}{\longrightarrow}  \ldots \overset{\alpha_n}{\longrightarrow}  s_{j} $) in the TS starting from $s_i$ and ending in $s_j$ with transitions labelled as $(\alpha_1 , ... , \alpha_n)$. 

To perform trace selection in a DTMC $dtmc$, we use a classical Depth First Search (DFS) algorithm parametrized with a maximum length $l_{max}$ for finite traces and an interval  $[Pr_{min} , Pr_{max}]$ specifying the minimal and maximal values for the probabilities of selected traces. Formally: 
\begin{align*}
    DFS & (l_{max} , Pr_{min}, Pr_{max}, dtmc) =
        \{(\alpha_1 , ... , \alpha_n ) \\ 
        & \mid i \overset{(\alpha_1 , ... , \alpha_n)}{\Longrightarrow} i \\
        & \wedge ( \not \exists k : 1 < k < n \wedge i \overset{(\alpha_1 , ... , \alpha_k)}{\Longrightarrow} i  ) \\
        & \wedge (Pr_{min} \leq Pr(\alpha_1 , ... , \alpha_n) \leq Pr_{max})
        \} 
\end{align*}
where $\tau_{dtmc} (i) = 1$ and $Pr(\alpha_1 , ... , \alpha_n) = \Pi_{k=1}^{n} P_{dtmc}(s_i , s_j)$ such as $s_i \overset{\alpha_k}{\longrightarrow}  s_{j} \in (i \overset{\alpha_1}{\rightarrow} \ldots \overset{\alpha_n}{\rightarrow} i)$.  We initially consider only finite traces starting from and ending in the initial state~$i$ (assimilate to an accepting state) without passing by~$i$ in between.  Finite sequences starting from and ending in~$i$ corresponds to a coherent execution scenario in the DTMC. With respect to partial finite traces (i.e., finite traces not ending in~$i$), our trace definition involve a smaller state space to explore in the DTMC. This is due to the fact that the exploration of a part of the graph may be stopped, without further checks of the existence of partial finite traces, as long as the partial trace is higher than~$l_{max}$. The DFS  algorithm is hence easier to implement and may better scale to large DTMCs.

Practically, this algorithm will build a n-tree where a node represents a state with the probability to reach it and the branches are the $\alpha_k$ labels of the transitions taken from the state associated to the node. The root node corresponds to the initial state~$i$ and has a probability of $1$. Since we are only interested in finite traces ending in the initial state, the exploration of a branch of the tree is stopped when the depth is higher than then maximal path~$l_{max}$. This parameter is provided to the algorithm by the test engineer and is only used to avoid infinite loops during the exploration of the DTMC. Its value will depend on the size of the DTMC and should be higher than the maximal ``loop free'' path in the DTMC in order to get coherent finite traces.

For instance, the execution of the algorithm on the soda vending machine ($vm$) example presented in Fig.~\ref{fig_FTS_vending_machine} gives 5 finite traces: 
\begin{align*}
& DFS  (7 ;  0 ; 0.1 ; DTMC_{vm}) = \{ \\ 
    & (pay, change, cancel, return) ; (free, cancel, return) ;\\
    & (pay, change, tea, serveTea, open, take,  close); \\
    & (pay, change, tea, serveTea, take) ; \\
    & (free, tea, serveTea, open, take, close)
    \}
\end{align*}
During the execution of the algorithm, the trace $(free, tea, serveTea, take)$ has been rejected since its probability ($0.729$) is is not between $0$ and $0.1$.

The downside is that the algorithm will possibly enumerate all the paths in the DTMC depending on the~$l_{max}$ value. This can be problematic and we plan in our future work to use symbolic executions techniques inspired by work in the probabilistic model checking area, especially automata-based representations~\cite{Baier2008} in order to avoid a complete state space exploration.

\subsubsection{Traces Filtering using the FTS and Building the FTS'}

\begin{figure}
\begin{algorithmic}[1]

    \REQUIRE $traces, fts$
    \ENSURE $traces, fts'$

    \STATE{$S_{fts'} \leftarrow \{i_{fts}\}$ ; $i_{fts'} \leftarrow i_{fts}$ ; $d_{fts'} \leftarrow d_{fts}$}    
    \FORALL{$t \in traces$}
        \IF{$ accept(fts, t) $}
            \STATE{$S_{fts'} \leftarrow S_{fts'} \cup states(fts, t)$}
            \STATE{$Act_{fts'} \leftarrow Act_{fts'} \cup t$}
            \STATE{$trans_{fts'} \leftarrow trans_{fts'} \cup transitions(fts, t)$}
            \STATE{$\gamma_{fts'} \leftarrow fLabels (fts, t) \gamma_{fts'}$}
        \ELSE
            \STATE{$traces \leftarrow traces \setminus \{t\}$}
        \ENDIF 
    \ENDFOR
    \RETURN $fts'$
    
\end{algorithmic}
    \caption{FTS' building algorithm}
    \label{algo_ftsprime_building}
\end{figure}

 Generated finite traces from the DTMC may contain illegal sequences of actions (i.e., sequences of actions which can not be performed by any valid product of the SPL). The set of generated finite traces has to be filtered using the FTS such that the following property holds: for a given FTS $fts$ and a usage model $dtmc$, a finite trace~$t$ generated from $dtmc$ represents a valid behaviour for the product line $pl$ modelled by $fts$ if there exists a product $p$ in $pl$ such as $t\subseteq  [\![fts_{\mid p}]\!]_{TS}$, where $fts_{\mid p}$ represents the projection of $fts$ using product $p$ and $[\![ts]\!]_{TS}$ represents all the possible traces and their prefixes for a TS $ts$. The idea here is to use the FTS to detect invalid finite traces by running them on it. 

Practically, we will build a second FTS' which will represent only the behavior of the SPL appearing in the finite traces generated from the DTMC. Fig. \ref{algo_ftsprime_building} presents the algorithm used to build an $fts'$ from a set of $traces$ (filtered during the algorithm) and a $fts$. The initial state of $fts'$ corresponds to the initial state of the $fts$ (line 1) and $d$ in $fts'$ is the same as for $fts$ (line 1). If a given trace is accepted by the $fts$ (line 3), then the states, actions and transitions visited in $fts$ when executing the trace $t$ are added to $fts'$ (line 4 to 6). 
The $accept(fts,t)$ function on line~3 will return true if there exists at least one product in $d_{fts}$ that can execute the sequence of actions in $t$.
On line~7, the $fLabels(fts, t)$ function is used to enrich the $\gamma_{fts'}$ function with the feature expressions of the transitions visited when executing $t$ on the $fts$. It has the following signature : $fLabels : (FTS, trace) \rightarrow \gamma \rightarrow \gamma$ and $fLabels (fts, t) \gamma_{fts'}$ will return a new function $\gamma'_{fts'}$ which will for a given transition $tr = (s_i \overset{\alpha_k}{\longrightarrow}  s_{j}) $ return $\gamma_{fts} tr$ if $\alpha_k \in t$ or $\gamma_{fts'} tr$ otherwise.

\begin{figure}[t]
\centering
\includegraphics[width=0.45\textwidth]{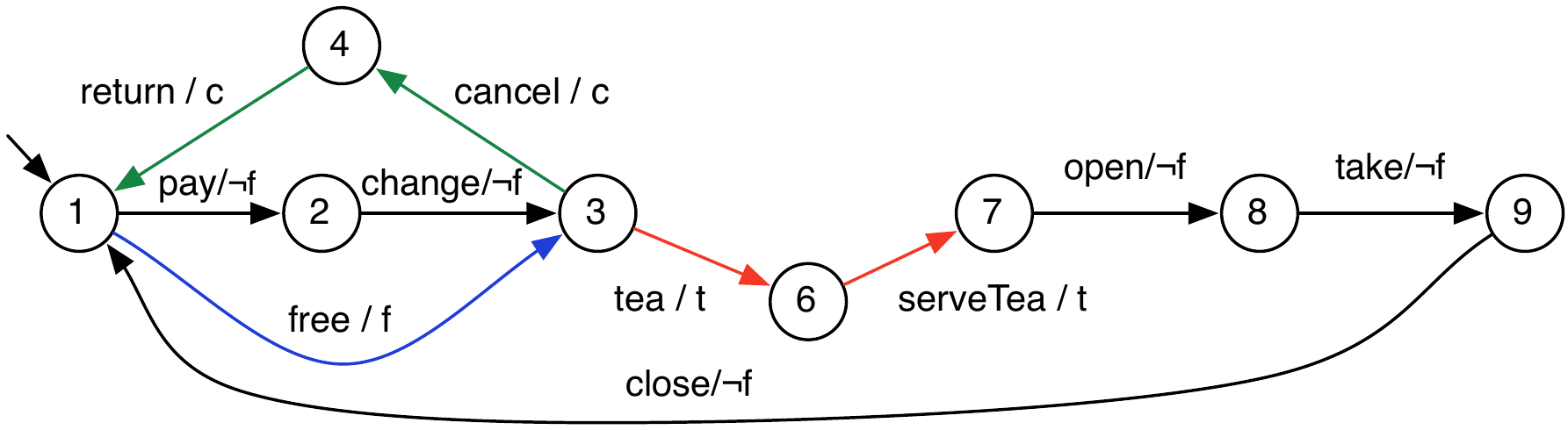}
\caption{FTS' of the soda vending machine}
\label{fig_fts_prime_vm}
\end{figure}

In our $vm$ example, the set of finite traces generated from step~1 contains two illegal traces: $(pay,$ $change,$ $tea,$ $serveTea,$ $take)$ and $(free,$ $tea,$ $serveTea,$ $open,$ $take,$ $close)$. Those 2 traces (mixing free and not free vending machines) cannot be executed on the $fts_{vm}$ and will be rejected in step~2. The generated $fts'_{vm}$ is presented in Fig.~\ref{fig_fts_prime_vm}.

\subsubsection{Product Prioritization}

At the end of step 2 in Fig.~\ref{fig_family_based_approach}, we have an FTS' and a set of finite traces in this FTS'. This set of finite traces (coming from the DTMC) covers all the valid behaviors of the FTS'. It is thus possible to order them according to their probability to happen. This probability corresponds to the the cumulated individual probabilities of the transitions fired when executing the finite trace in the DTMC. 
A valid finite trace $t=(\alpha_1 , \ldots , \alpha_n)$ corresponding to a path $(s_1 \overset{\alpha_1}{\longrightarrow} \ldots \overset{\alpha_n}{\longrightarrow} s_{n+1})$ in the DTMC (and in the FTS') has a probability $Pr(t)$ (calculated as in step~1) to be executed. We may perform bookkeeping of $Pr(t)$. 

The set of products able of executing a trace $t$ may be calculated from the FTS' (and its associated FD). It corresponds to all the products (i.e., set of features) of the FD ($[\![d]\!]$) that satisfy all the feature expressions associated to the transitions of $t$. Formally, for $t$ and a FTS' $fts'$, the set of products~$ prod(t,fts') = \bigcap_{k=1}^{n} \{p \mid \gamma_{fts'} (s_k \overset{\alpha_k}{\longrightarrow} s_{k+1}) p = true \}$. 
From a practical point of view, the set of products corresponds to the products satisfying the conjunction of the feature expressions $\gamma_{fts'} (s_k \overset{\alpha_k}{\longrightarrow} s_{k+1}) $ on the path of  $t$ and the FD~$d_{fts'}$. As $d_{fts'}$ may be transformed to a boolean formula where features become variables~\cite{Czarnecki2007}, the following formula can be calculated using a SAT solver:~$\bigwedge_{k=1}^{n} (\gamma_{fts'} (s_k \overset{\alpha_k}{\longrightarrow} s_{k+1})) \wedge booleanForm(d_{fts'})$.

At this step, each valid finite trace $t$ is associated to the set of products $ prod(t,fts') $ that can actually execute~$t$ with a probability~$Pr(t)$. Product priortization may be done by classifying the finite traces according to their probability to be executed, giving $t$-behaviorally equivalent classes of products for each finite trace $t$. For instance, for the trace $t_{vm}=(pay, change, tea, serveTea, open, take,  close)$ generated for our $vm$ example the products will have to satisfy the $\neg f \wedge t $ feature expression and $d_{vm}$. This gives us a set of 8 products (amongst 32 possible):
\begin{align*}
\{&(v, b, cur, t, eur) ; (v, b, cur, t, usd) ; (v, b, cur, t, c, eur) ; \\
& (v, b, cur, t, c, usd) ; (v, b, cur, t, s, eur) ; (v, b, cur, t, s, usd) ;\\
&(v, b, cur, t, s, c, eur) ; (v, b, cur, t, s, c, usd) \}
\end{align*}
All of them executing $t_{vm}$ with a probability $Pr(t_{vm}) = 0.009$ which is the lowest probable behaviour of the soda vending machine.

\subsubsection{Model Checking and Test Case Generation}

Since the FTS' represents the set of valid products capable of executing the valid finite traces generated from the DTMC. It represents a subset of the behavior of the SPL that has to be assessed  in priority according to the provided $[Pr_{min} , Pr_{max}]$ bounds. The FTS' may be verified using existing algorithms \cite{Classen2013} and/or be used to generate test cases~\cite{Devroey2012}.

\section{Related Work} \label{relatedWork}

To the best of our knowledge, there is no approach prioritizing behaviors statistically for testing SPLs in a family-based manner.  The most related proposal (outlined in section \ref{approach}) has been devised by Samih and Baudry \cite{Samih2012}.  This is a product-based approach and therefore requires selecting one or more products to test at the beginning of the method. One also needs that the DTMC covers all products of the SPL, which is not our assumption here. 

There have been SPL test efforts to sample products for testing such as t-wise approaches (e.g.  \cite{PerrouinOSKBT12,Cohen2006,Cohen2007}). More recently sampling was combined with prioritization thanks to the addition of weights on feature models and the definition of multiple objectives \cite{DBLP:conf/models/JohansenHFES12,splc13}.  However, these approaches do not consider SPL behavior in their analyses.  

To consider behavior in an abstract way, a full-fledged MBT approach \cite{utting2007practical} is required. Although behavioural MBT is well established for single-system testing~\cite{Tretmans2008MBT-with-LTS}, a survey  \cite{Oster2011MBSPL-testing-survey} shows insufficient support of SPL-based MBT. However, there have been efforts to combine sampling techniques with modeling ones (e.g. \cite{Lochau2012-paiwise-sc}).  These approaches are also product-based, meaning that may miss opportunities for test reuse amongst sampled products \cite{DBLP:conf/vamos/RheinAKTS13}. We believe that benefiting from the recent advances in behavioral modeling provided by the model checking community \cite{Asirelli2011,Asirelli2011a,classen_symbolic_2011,classen_model_2010,Fischbein-2006-Found-behavioural-conformance-in-SPL-archi,barthe_modeling_2008,lauenroth_model_2009,Baier2008}, sound MBT approaches for SPL can be derived and interesting family-based scenarios combining verification and testing can be devised \cite{Devroey2012}.

Our will is to apply ideas stemming from statistical testing and adapt them in an SPL context. For example, combining structural criteria with statistical testing has been discussed in \cite{Gouraud:2001:NWA:872023.872550,DBLP:journals/stvr/Thevenod-FosseW91}. We do not make any assumption on the way the DTMC is obtained: via an operational profile \cite{musa1996operational} or by analyzing the source code or the specification \cite{DBLP:journals/stvr/Thevenod-FosseW91}. However, an uniform distribution of probabilities over the DTMC would probably be less interesting. As noted by Witthaker \cite{Whittaker:1994fk}, in such case only the structure of traces would be considered and therefore basing their selection on their probabilities would just be a means to limit their number in a mainly random-testing approach. In such cases, structural test generation has to be employed \cite{feliachi2010generating}.         

\section{Conclusion} \label{conclusion}

In this paper, we combine concepts stemming from statistical testing with SPL sampling to extract products of interest according to the probability of their execution traces gathered in a discrete-time markov chain representing their usages. As opposed to product-based sampling approaches, we select a subset of the full SPL behavior given as Featured Transition Systems (FTS).  This allows us to construct a new FTS representing only the executions of relevant products.  This such pruned FTS can be analyzed all at once, to enable allow test reuse amongst products and/or to scale model-checking techniques for testing and verification activities.
Future work will naturally proceed to the full implementation of the approach presented here and its validation on concrete systems.  This raise a number of challenges, including inference of usage models using various techniques such as the analysis of systems logs or symbolic execution of the software product line, as well as the design of efficient algorithms for trace extraction and FTS pruning. We also would like to consider partial traces (traces that do not need to end in the initial state). Although making prioritization less scalable, they may prove useful when the discrepancies between the behavioral and usage models are too important (partial execution can cope such situations easily) or to focus on specific feature interactions.

\bibliographystyle{IEEEtran}
\bibliography{StatisticalPrioritizationSPL}
%
%
%

\end{document}